# Radiation and Dynamics of Nanoparticle in Equilibrium Background Radiation upon Translational-Rotational Motion

G. V. Dedkov[*] and A. A. Kyasov

Kabardino-Balkarian State University, Nanoscale Physics Group, Nalchik, Russia
[*]e-mail: gv_dedkov@mail.ru

We have obtained general expressions for the intensity of radiation and tangential force of a small polarizable particle in the process of translational–rotational motion in equilibrium radiation background (thermalized photon gas) of certain temperature at an arbitrary relative orientation of the linear and angular velocity vectors. It is shown that in cold vacuum background the translational velocity of particle is independent of time and intensity of its spontaneous emission is determined by the angular velocity and imaginary part of the particle polarizability.

Excessive electromagnetic radiation (superradiance) of rotating cylinder upon scattering of incident photons was first noted by Zel'dovich [1]. Quite recently, this issue has been developed obtained further when studying radiation of rotating and moving neutral polarized bodies in the vacuum background of zero temperature and in an equilibrium background radiation (thermalized photon gas) with certain temperature [2—5]. In contrast to [2—5], this work aims at studying the general case of relativistic translational-rotational motion of small particle in radiation background assuming arbitrary relative orientation of angular and linear velocity vectors.

We assume that a small spherical particle of radius $R$ and temperature $T_1$ (in the own rest frame $\Sigma''$) moves with linear velocity $\mathbf{V}$ relative to radiation background (reference frame $\Sigma$) and rotates with angular velocity $\mathbf{\Omega} = \Omega\mathbf{n}$ in comoving reference frame $\Sigma'$, $\theta$ is the angle between $\mathbf{\Omega}$ and $\mathbf{V}$ vectors (see Fig. 1). We also assume that an equilibrium vacuum background has temperature $T_2$ and the following conditions are satisfied: $\Omega R/c << 1$, $R << \min(2\pi\hbar c/k_B T_1, \ 2\pi\hbar c/k_B T_2)$, $\hbar, k_B, c$ – are the Planck's, Boltzmann's constants and the speed of light in vacuum. In this case, when emitting low-frequency photons, a particle can be treated as a point-like dipole with fluctuating electric and magnetic dipole moments $\mathbf{d}(t), \mathbf{m}(t)$. The material properties of the particle are described by frequency-dependent dielectric $\alpha_e(\omega)$ and magnetic $\alpha_m(\omega)$ polarizabilities defined in frame of reference $\Sigma$ of the particle.

Let a rather distant surface $\sigma$ surrounds the particle so that a fluctuating electromagnetic field on $\sigma$ has a wave character. According to the law of conservation of energy of the system in volume (V) restricted by $\sigma$ one can write

$$-\frac{dW}{dt} = \oint_\sigma \mathbf{S} \cdot d\vec{\sigma} + \int_{(V)} \langle \mathbf{j} \cdot \mathbf{E} \rangle d^3 r, \qquad (1)$$

where $W = (1/8\pi) \int_{(V)} (\langle \mathbf{E}^2 \rangle + \langle \mathbf{H}^2 \rangle) d^3 r$ is the energy of fluctuating field with components **E** и **H** in volume (V), $\mathbf{S} = (c/4\pi)\langle \mathbf{E} \times \mathbf{H} \rangle$ is the Pointing vector, and **j** the current density. In the quasi-stationary approximation, $dW/dt = 0$, from (1) it follows that

$$I = \oint_\sigma \mathbf{S} \cdot d\vec{\sigma} = -\int_{(V)} \langle \mathbf{j} \cdot \mathbf{E} \rangle d^3 r \equiv I_1 - I_2 . \qquad (2)$$

Here $I_1 = I_1(T_1)$ is the intensity of radiation emitted by the particle in vacuum, and $I_2 = I_2(T_2)$ is the intensity of absorbed background radiation. Using relativistic transformations of **j, E** and volume (V), formula (2) takes a simpler form [4, 5]

$$I = I_1 - I_2 = -\left(\frac{dQ}{dt} + F_x V\right), \qquad (3)$$

$$dQ/dt = \langle \mathbf{\dot{d}} \cdot \mathbf{E} + \mathbf{\dot{m}} \cdot \mathbf{H} \rangle, \qquad (4)$$

$$F_x = \langle \nabla(\mathbf{d} \cdot \mathbf{E} + \mathbf{m} \cdot \mathbf{H}) \rangle_x . \qquad (5)$$

All quantities in (1)—(5) correspond to frame of reference $\Sigma$ of the vacuum background, angular brackets indicate the full quantum-statistical averaging, and the points above **d** and **m** mean the time differentiation. The $F_x$ in (3) and (5) describes the tangential force applied to the particle, and $dQ/dt$ defines the rate of particle heating. The calculations in (4), (5) are carried out within a general approach of fluctuating electrodynamics [6] developed for the problems with dynamic and thermal disequilibrium. In this case, the right-hand sides of (4) and (5) are represented in the form of pairwise products of spontaneous and induced components of the fields and dipole moments. In contrast to [5], to transform the components of vectors given in reference frame $\Sigma'$ and in rotating system $\Sigma''$, the general form of the rotation matrix is used

$$A_{ik} = n_i n_k + (\delta_{ik} - n_i n_k)\cos\Omega\tau - e_{ikl} n_l \sin\Omega\tau, \qquad (6)$$

where $n_i$ are the components of the unit vector **n** of rotation axis in Cartesian coordinates of system $\Sigma'$, $\tau$ is the time in frame $\Sigma''$, and $e_{ikl}$ is the unit antisymmetric tensor of the third rank. Due to the condition $\Omega R/c \ll 1$, the variables of time $\tau$ and $t'$ (time in $\Sigma'$) are the same. With allowance for (6), required fluctuation-dissipation relationships for the components of spontaneous dipole moment of particle in reference frame $\Sigma'$ take the form

$$\left\langle d^{sp}{}_x{}'(\omega)d^{sp}{}_x{}'(\omega')\right\rangle = \frac{1}{2}2\pi\hbar\delta(\omega+\omega')\cdot$$
$$\left\{2\cos^2\theta\,\alpha''_e(\omega)\coth\frac{\hbar\omega}{2k_BT_1} + \sin^2\theta\left[\alpha''_e(\omega^+)\coth\frac{\hbar\omega^+}{2k_BT_1} + \alpha''_e(\omega^-)\coth\frac{\hbar\omega^-}{2k_BT_1}\right]\right\}, \quad (7)$$

$$\left\langle d^{sp}{}_y{}'(\omega)d^{sp}{}_y{}'(\omega')\right\rangle = \frac{1}{2}2\pi\hbar\delta(\omega+\omega')\cdot$$
$$\left\{2\sin^2\theta\,\alpha''_e(\omega)\coth\frac{\hbar\omega}{2k_BT_1} + \cos^2\theta\left[\alpha''_e(\omega^+)\coth\frac{\hbar\omega^+}{2k_BT_1} + \alpha''_e(\omega^-)\coth\frac{\hbar\omega^-}{2k_BT_1}\right]\right\}, \quad (8)$$

$$\left\langle d^{sp}{}_z{}'(\omega)d^{sp}{}_z{}'(\omega')\right\rangle = \frac{1}{2}2\pi\hbar\delta(\omega+\omega')\cdot\left[\alpha''_e(\omega^+)\coth\frac{\hbar\omega^+}{2k_BT_1} + \alpha''_e(\omega^-)\coth\frac{\hbar\omega^-}{2k_BT_1}\right], \quad (9)$$

where $\omega^\pm = \omega \pm \Omega$. Analogous relations for spontaneous magnetic moment of particle are obtained from (7)—(9) replacing $\alpha_e(\omega)$ by $\alpha_m(\omega)$.

Using (7)—(9) along with the magnetic analogs and calculating the right-hand sides in (4), (5) as in [4, 5], one obtains

$$F_x = -\frac{\hbar\gamma}{4\pi c^4}\int_{-\infty}^{+\infty}d\omega\,\omega^4\int_{-1}^{1}dx\,x\cdot$$
$$\left\{\begin{array}{l}\left[(1-\beta^2)(1-x^2)\cos^2\theta + \left((1+\beta^2)(1+x^2)+4\beta x\right)\dfrac{\sin^2\theta}{2}\right]\cdot \\ \alpha''(\gamma\omega(1+\beta x))\left(\coth\dfrac{\hbar\omega}{2k_BT_2} - \coth\dfrac{\hbar\gamma\omega(1+\beta x)}{2k_BT_1}\right) + \\ +\left[(1-\beta^2)(1-x^2)\sin^2\theta + \left((1+\beta^2)(1+x^2)+4\beta x\right)\dfrac{1+\cos^2\theta}{2}\right]\cdot \\ \alpha''(\gamma\omega(1+\beta x)+\Omega)\left(\coth\dfrac{\hbar\omega}{2k_BT_2} - \coth\dfrac{\hbar(\gamma\omega(1+\beta x)+\Omega)}{2k_BT_1}\right)\end{array}\right\}, \quad (10)$$

$$\dot{Q} = \frac{\hbar \gamma}{4\pi c^3} \int_{-\infty}^{+\infty} d\omega \omega^4 \int_{-1}^{1} dx (1+\beta x) \cdot$$

$$\cdot \left\{ \begin{array}{l} \left[ (1-\beta^2)(1-x^2)\cos^2\theta + \left((1+\beta^2)(1+x^2)+4\beta x\right)\frac{\sin^2\theta}{2} \right] \cdot \\ \alpha''(\gamma\omega(1+\beta x))\left( \coth\frac{\hbar\omega}{2k_B T_2} - \coth\frac{\hbar\gamma\omega(1+\beta x)}{2k_B T_1} \right) + \\ + \left[ (1-\beta^2)(1-x^2)\sin^2\theta + \left((1+\beta^2)(1+x^2)+4\beta x\right)\frac{1+\cos^2\theta}{2} \right] \cdot \\ \alpha''(\gamma\omega(1+\beta x)+\Omega)\left( \coth\frac{\hbar\omega}{2k_B T_2} - \coth\frac{\hbar(\gamma\omega(1+\beta x)+\Omega)}{2k_B T_1} \right) \end{array} \right\}, \quad (11)$$

where $\beta = V/c, \gamma = (1-\beta^2)^{-1/2}$, and $\alpha'' = \alpha_e'' + \alpha_m''$ – is a sum of the imaginary parts of the electric and magnetic polarizabilities. Substituting (10), (11) into (3) yields

$$I = I_1 - I_2 = \frac{\hbar\gamma}{4\pi c^3} \int_{-\infty}^{+\infty} d\omega \omega^4 \int_{-1}^{1} dx \cdot$$

$$\cdot \left\{ \begin{array}{l} \left[ (1-\beta^2)(1-x^2)\cos^2\theta + \left((1+\beta^2)(1+x^2)+4\beta x\right)\frac{\sin^2\theta}{2} \right] \cdot \\ \alpha''(\gamma\omega(1+\beta x))\left( \coth\frac{\hbar\gamma\omega(1+\beta x)}{2k_B T_1} - \coth\frac{\hbar\omega}{2k_B T_2} \right) + \\ + \left[ (1-\beta^2)(1-x^2)\sin^2\theta + \left((1+\beta^2)(1+x^2)+4\beta x\right)\frac{1+\cos^2\theta}{2} \right] \cdot \\ \alpha''(\gamma\omega(1+\beta x)+\Omega)\left( \coth\frac{\hbar(\gamma\omega(1+\beta x)+\Omega)}{2k_B T_1} - \coth\frac{\hbar\omega}{2k_B T_2} \right) \end{array} \right\}. \quad (12)$$

From (12) one can get all particular results previously obtained in [4, 5, 7]. The most important consequence of general formulas (10), (12) is the possibility of non-thermal emission in the system "vacuum-rotating particle". Thus, at $T_1 = T_2 = 0$ from (12) and (10) one obtains

$$I^{(0)} \equiv I_1(0) = \frac{\hbar\gamma}{2\pi c^3} \int_{-1}^{1} dx \cdot f(x,\theta) \int_{0}^{\Omega\gamma^{-1}(1+\beta x)^{-1}} d\omega \omega^4 \alpha''(\Omega - \gamma\omega(1+\beta x)) = $$

$$= \frac{4\hbar}{3\pi c^3} \int_{0}^{\Omega} d\xi \xi^4 [\alpha_e''(\Omega-\xi) + \alpha_m''(\Omega-\xi)], \quad (13)$$

$$f(x,\theta) = (1-\beta^2)(1-x^2)\sin^2\theta + \left((1+\beta^2)(1+x^2)+4\beta x\right)\frac{1+\cos^2\theta}{2}, \quad (14)$$

$$F_x^{(0)} = -\frac{4\hbar V}{3\pi c^5} \int_0^\Omega d\xi \xi^4 \left[\alpha_e''(\Omega-\xi) + \alpha_m''(\Omega-\xi)\right] \tag{15}$$

As we can see from the right-hand side of (13), integral intensity of radiation upon translational-rotational motion of particle does not depend neither on the velocity $V$ nor on the angle $\theta$. In contrast to that, the spectral and angular intensity distributions (see (13) and (14)) considerably depend on $V$ и and $\theta$. From (13) and (15) we directly obtain the relation

$$F_x = -\frac{\beta}{c} \cdot I^{(0)}. \tag{16}$$

The issue of particle dynamics should be considered in more detail. For this purpose, we rewrite the energy conservation law (1) in the form

$$-\frac{d}{dt}\left(W + \frac{mc^2}{\sqrt{1-\beta^2}}\right) = I_1 - I_2 = I, \tag{17}$$

and use the dynamics equation

$$\frac{d}{dt}\left(\frac{mc\beta}{\sqrt{1-\beta^2}}\right) = F_x, \tag{18}$$

where $m$ is the particle mass. In (17) and (18), the $m$ and $\beta$ depend on time, while $dW/dt = 0$ due to the condition of quasistationarity, as in (1). By eliminating derivative $dm/dt$ from (18) with the help of (17) and taking into account (3), we rewrite Eq. (18) in the form

$$\gamma^3 mc \frac{d\beta}{dt} = F_x - \frac{\beta}{1-\beta^2}\frac{\dot{Q}}{c}. \tag{19}$$

Equation (19) is valid at arbitrary temperatures $T_1$ and $T_2$. In particular, at $T_1 = T_2 = 0$, with allowance for (3) and (16), from (19) one obtains $d\beta/dt \equiv 0$.

Thus, particle rotation and its spontaneous non-thermal radiation do not affect the velocity of translational motion. At a finite temperature $T_2$ of the vacuum background, the issue of radiation, dynamics of translational-rotational motion and kinetics of particle heating needs special investigation. In the absence of rotation, as shown in [4], the particle dynamics depends only on the temperature of radiation background. In the rest frame of particle, as time elapses, the particle temperature takes on an effective value depending on $T_2$ and $\gamma$, and the intensity of thermal emission is determined by this effective temperature [4].

The results should be interesting for astrophysics of cosmic dust matter and particle dynamics in traps, as well as in creating the sources of directional microwave and radio emission. The effect of directional thermal radiation of moving particles may affect an observed anisotropy of the primary radiation background.

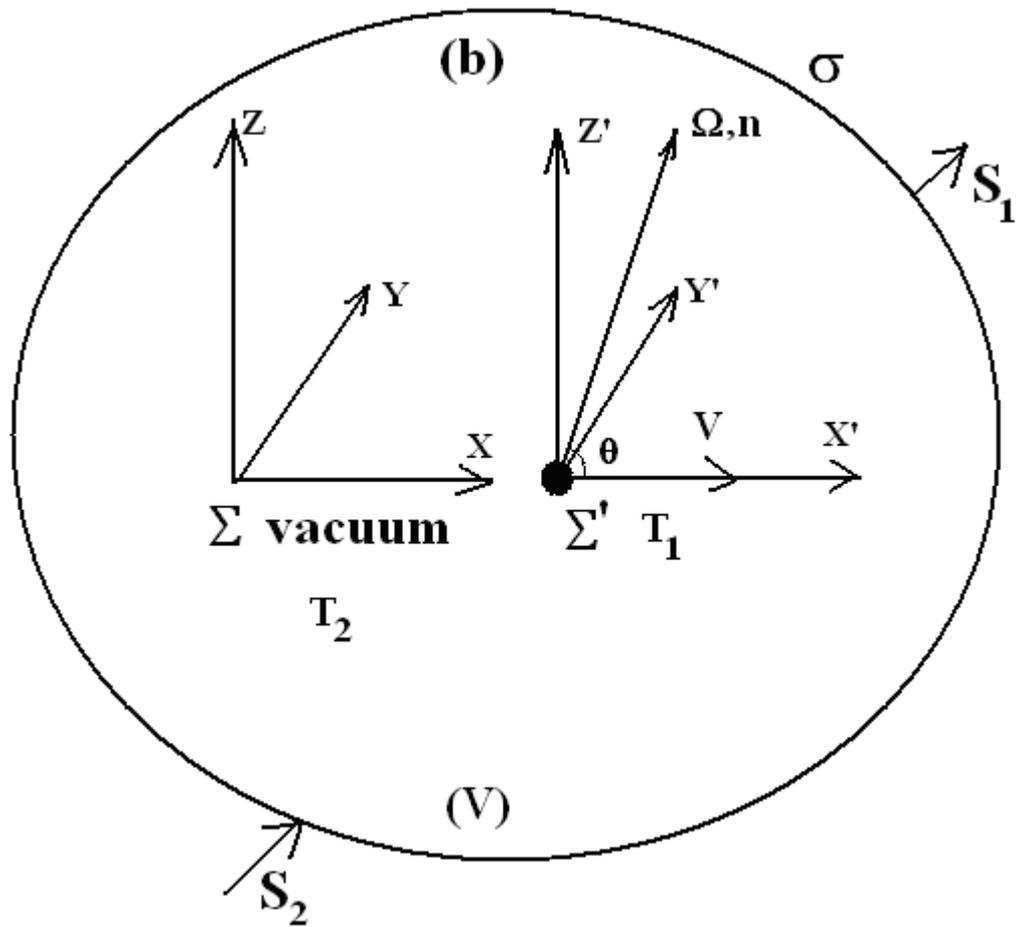

Figure. Schematics of particle motion and coordinate systems used. The coordinate system $\Sigma''$ corresponding to the rest frame of particle is not shown. Its axis $Z''$ is directed along the vector $\Omega$.